\documentclass[fleqn,10pt,twocolumn]{SICE14_fixed}

\usepackage{amsmath,fleqn,bm}

\newcommand{\bea}{\begin{eqnarray}}
\newcommand{\eea}{\end{eqnarray}}
\newcommand{\ba}{\begin{array}}
\newcommand{\ea}{\end{array}}

\newcommand{\be}{\begin{equation}}

\newcommand{\ee}{\end{equation}}
\newcommand{\bt}{\begin{teo}}
\newcommand{\et}{\end{teo}}

\newcommand{\imag}{{\rm i}}

\title{Accelerator modes and their effect in the diffusion properties in the kicked rotator}

\author{Thanos Manos${}^{\dag,1,2,3}$ and Marko Robnik${}^{4}$}
\speaker{Thanos Manos}

\affils{${}^{1}$CAMTP - Center for Applied Mathematics and Theoretical Physics, University of Maribor, Krekova 2, SI-2000 Maribor, Slovenia\\
${}^{2}$School of Applied Sciences, University of Nova Gorica, Vipavska 11c, SI-5270 Ajdov\v s\v cina, Slovenia\\
${}^{3}$ Institute of Neuroscience and Medicine (INM-7), Research Center J\"{u}lich, Germany\\
(E-mail: thanos.manos@uni-mb.si,  t.manos@fz-juelich.de)\\
${}^{4}$CAMTP - Center for Applied Mathematics and Theoretical Physics, University of Maribor, Krekova 2, SI-2000 Maribor, Slovenia\\
(E-mail: Robnik@uni-mb.si)\\
}

\abstract{We highlight a few recent results on the effect of the diffusion
process in deterministic area preserving maps with noncompact phase space,
namely the standard map. In more detail, we focus on the anomalous diffusion
arising due to the accelerator modes, i.e. resonant-like features of the phase
space which are transported in a super-diffusive (ballistic) manner. Their
presence affects also the trajectories lying in the immediate neighborhood
resulting in anomalous (non-Gaussian) diffusion. We aim to shed some light on
these special properties of the phase space by utilizing the diffusion exponent (rate of diffusion) and the momentum distribution in terms of the L\'evy stable distributions, with the goal to reach an understanding of the global behaviour. To this end we consider a rather large interval for the kick parameter ($0\le K \le 70$) of the standard map where accelerator modes of different periodicity exist.}

\begin{document}

\maketitle


\section{Introduction}

In this paper we present a small review of our recent findings on the diffusion properties in area preserving maps, exemplified by the standard map
\cite{Chir1979}. This work has been linked to the extensive studies in
\cite{MR2013,BMR2013} of the quantum kicked rotator (see e.g. \cite{IZ1990}),
where the classical diffusion process also plays a significant role in the
quantum semiclassical level, where the (exponential) quantum (or dynamical)
localization is observed and studied. We here present a few results on how the presence of accelerator modes affects the dynamics of the classical phase space, and in more detail its diffusion rates and transport. We pursue this goal by studying the diffusion process of ensembles of initial conditions for a rather large value-interval of the control (kick) parameter ($0\le K \le 70$). Within this interval the phase space of the model shows a very rich variety of dynamics, namely from the almost entirely regular phase space to gradually more and more chaotic dynamics until there are practically no stable regions
anymore. In this picture there are some resonance features, the so-called
accelerator modes, that may affect the diffusion of trajectories rather
drastically.

Here, we are trying to shed some light on these effects by calculating the
\textit{diffusion exponent} $\mu$, the corresponding \textit{diffusion
constant} $D_{\mu}$ and the parameters of the relevant \textit{L\'evy stable
distribution} in the case of non-Gaussian (anomalous) diffusion for different
$K$ parameter values. Comparing to previous studies, here and mainly in
\cite{MR2014} we do not restrict ourselves to local effects and isolated
accelerator modes but we are interested in a rather more global point of view.
Namely, on how these features may influence the transport in their
neighborhood. That is the case when one, for example, considers mixed ensembles of initial conditions, e.g. ensembles that contain accelerator modes together
with chaotic trajectories around them. Let us also stress here that the time
scales we are interested in in this particular study are relatively short and we are not interested at this point on the asymptotic diffusion rates. This has to do with our initial motivation in understanding the quantum effects
associated with such features, namely  the  quantum localization properties
(see \cite{MR2013,BMR2013} for more details).

The paper is structured as follows: In Sec.~\ref{sec:model} we give brief
description of the model (standard map), while in Sec.~\ref{sec:results} we
present our main results on the dynamical effect of accelerator modes on the
diffusion exponent and the stable L\'evy distribution parameters. Finally, in
Sec.~\ref{sec:summary} we discuss the main findings of this work.

\section{The model} \label{sec:model}

For the purposes of our study we use a rather well-known time-dependent system, the paradigm of classical chaos in Hamiltonian (nondissipative) systems,
namely the kicked rotator \cite{Chir1979}. Its Hamiltonian function is given by
\be \label{KR} H= \frac{p^2}{2I} + V_0 \,\delta_T(t)\,\cos \theta. \ee  Here
$p$ is the (angular) momentum, $I$ the moment of inertia, $V_0$ is the strength of the periodic kicking, $\theta$ is the (canonically conjugate, rotation)
angle, and $\delta_T(t)$ is the periodic Dirac delta function with period $T$.
Taking into account that between the kicks the rotation is free, the
Hamiltonian equations of motion can be easily integrated, and thus the dynamics can be reduced to the standard mapping, or so-called Chirikov-Taylor mapping,
given by \be \label{SM1}
 \left\{
 \begin{aligned} \label{SM1}
  p_{n+1} &= p_n + V_0 \sin \theta_{n+1},\\
 \theta_{n+1} &= \theta_n + \frac{T}{I} p_n,
 \end{aligned}
 \right.
\ee
Note that the quantities $(\theta_n, p_n)$ refer to their values just
immediately after the $n$-th kick. Then, by introducing new dimensionless
momentum $P_n = p_nT/I$, we get \be
 \left\{
 \begin{aligned} \label{SM2}
P_{n+1} &= P_n + K \sin \theta_{n+1},\\
\theta_{n+1} &= \theta_n  + P_n,
 \end{aligned}
 \right.
\ee
where the system is now governed by a single classical {\em dimensionless}
kick parameter $K=V_0 T/I$, and the mapping is area preserving.

The generalized diffusion process of the standard map [Eq.~(\ref{SM2})] is then
defined as \be \label{varp} \langle(\Delta P)^2\rangle = D_{\mu}(K) n^{\mu},
\ee where $n$ is the number of iterations (kicks), and the exponent $\mu$ is in the interval $[0,2)$, and all variables $P$, $\theta$ and $K$ are
dimensionless. By $D_{\mu}(K)$ we define the generalized classical diffusion
constant and then according to the $\mu$-value one may distinguish further the
different types (rates) of diffusion. In more detail: in the case $\mu=1$ we
have the \textit{normal diffusion}, and $D_1(K)$ is then the normal diffusion
constant, while in the case of anomalous diffusion we observe \textit{subdiffusion} when $0 < \mu < 1$ or \textit{superdiffusion} if $1 <\mu
\le2$. In the case $\mu=2$ we have the \textit{ballistic transport} which is
associated strictly - as we found -  with the presence of accelerator modes.

Whenever the diffusion is normal ($\mu=1$) the theoretical value of $D_1(K)$ is
found to be \cite{IZ1990}
\begin{flalign} \label{Dcl}
 D_{1}(K)=
\begin{cases}
 \frac{1}{2} K^2\left [1- 2J_2(K) \left (1-J_2(K) \right ) \right ], \text{if} \ K \ge 4.5 \\
 0.15(K-K_{cr})^3, \text{if} \ K_{cr} < K \le 4.5
\end{cases}\hspace{-0.5cm},
\end{flalign}
where $K_{crit}\simeq 0.9716$ and $J_2(K)$ is the Bessel function (the higher
terms of order $K^{-2}$ are here neglected).

The dependence of the diffusion constant for the growth of the variance of the
momentum on $K$ is not trivial and fails to be in a good agreement around
the period 1 accelerator mode intervals
\be \label{acmdint} (2\pi n) \le K \le
\sqrt{(2\pi n)^2 +16 },
\ee
where $n$ is any positive integer (see \cite{IZ1990,MR2014}). In more detail, in these intervals for the accelerator modes $n=1$ there are two \textit{stable fixed points} located at $p=0,\; \theta = \pi -\theta_0$ and $p=0,\; \theta = \pi +\theta_0$, where $\theta_0 =
\arcsin (2\pi/K)$. There are also two \textit{unstable fixed points} at $p=0,\;
\theta = \theta_0$ and $p=0,\; \theta = 2\pi - \theta_0$.  Furthermore, as the
diffusion might even be anomalous, we have recalculated the \textit{effective}
diffusion constant $D_{\rm eff}=\langle(\Delta P)^2\rangle/n$ numerically,
which in general is not equal to the $D_{\mu}$ defined in Eq.~(\ref{varp})
(more detail about this process can be found in \cite{MR2014}).

\section{Results} \label{sec:results}

\begin{figure}[h]
\begin{center}
\includegraphics[width=8cm]{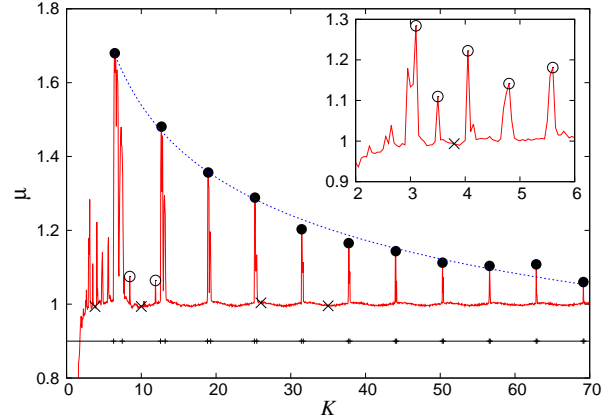}
\caption{\label{figKvsmu} (Color online) The diffusion exponent $\mu$ vs. $K$
for an ensemble of $\approx$100000 $(314\times 314)$ initial conditions
uniformly spread on the plane $(\theta,P)=(0,2\pi)$. The intervals of stable
accelerator modes of period 1 are depicted on the black horizontal line
$\mu=0.9$ while all the intervals of $K$ where the diffusion is normal are
found to have a diffusion exponent $\mu \approx 1$). Note that the larger
relatively peaks (for $K > 2\pi$ - black full circles) correspond accelerator
modes (mainly of period 1). Those peaks of smaller amplitude and for $K < 2\pi$ (enlarged in the inset) as well as those for $2\pi < K < 4\pi$ marked with
empty circle are related to accelerator modes of higher period. There are few
$K$ values marked with the symbol ($\times$) associated with normal diffusion.
All these cases where thoroughly studied in \cite{MR2014} where the interested
reader may find more details. The power law describing the decay of the
exponent $\mu$ of the main peaks' amplitude due to accelerator modes of period
1 (see text for more details) are shown by the blue dotted line.}
\end{center}
\end{figure}

We have calculated the diffusion exponent $\mu$ as a function of $K$. The final number of iterations is here $n=5000$ and the total number of initial
conditions is $(\approx 100000)$ spread uniformly on a $314\times 314$ grid on
the plane $(\theta,P)=(0,2\pi)$ (Fig.~\ref{figKvsmu}).
The diffusion exponent $\mu$ was calculated as described in Sec.~II
[Eq.~(\ref{varp})] in \cite{MR2014}, i.e. by the slopes of the lines of the
variance of the momentum $P$ as a function of iterations. The black horizontal
line $\mu=0.9$ indicates the intervals of stable accelerator modes of period 1
[Eq.~(\ref{acmdint})]. All intervals of $K$ associated with normal diffusion
processes have $\mu \approx 1$. The large peaks, marked with full
black circles, that appear for $K > 2\pi$ are due to the anomalous diffusion
raised by the accelerator modes of period 1. On the other hand, there are
a few relatively smaller peaks for $K < 2\pi$ (see the inset panel in
Fig.~\ref{figKvsmu}), whose origin is accelerator modes of higher period, and
also for $2\pi < K < 4\pi$, both these sets are marked with empty circles. With the symbol ($\times$) we mark few typical examples, close to those peaks, for
which the diffusion is normal. All these cases where thoroughly studied in
\cite{MR2014} where the interested reader may find more details.

In \cite{MR2014} it was found that the size of the large peaks for $K >
2\pi$ (full black circles in Fig.~\ref{figKvsmu} associated with regimes with
accelerator modes of period 1) decays monotonically with a power law \be f(x)=cx^d, \ee where $c=2.41645$ and $d=-0.195896$ [blue dotted line in
Fig.~\ref{figKvsmu}, with asymptotic standard error $\pm$0.04294 (1.777$\%$)
and $\pm$0.00537 (2.741$\%$) respectively]. Hence, this shows that their effect decreases significantly for large $K$-values, becoming probably entirely
insignificant for $K>70$. Moreover, the size of the successive accelerator modes of period 1 intervals decays with a power law defined simply and analytically by the Eq.~(\ref{acmdint}).

\subsection{The L\'evy stable distribution}\label{sec:Levy}
Let us here give a brief description of the L\'evy stable distribution. For a
random variable $X$ with distribution function $F(x)$, the characteristic
function is defined by $\phi(u)=E \exp(\imag u X)=\int_{-\infty}^{\infty}
\exp(\imag u x) dF(x)$. Then, a random variable $X$ is \textit{stable} if and
only if $X\overset{\delta}{=}aZ+b$, with $a>0$, $b \in \bm{R}$ and $Z$ is a
random variable with characteristic function
\begin{flalign} \label{Zcf}
 E \exp(\imag u Z) =
  \begin{cases}
  e^{-|u|^{\alpha}[1-\imag \beta \tan \frac{\pi \alpha}{2}({\rm sign}u)]},  \ & \alpha \neq 1 \\
  e^{-|u|[1+\imag \beta \tan \frac{2}{\pi}({\rm sign}u)] \log|u|}, \ & \alpha = 1 \end{cases},
\end{flalign}
where $0 < \alpha \le 2$ and $-1 \le \beta \le 1$ (the symbol
$\overset{\delta}{=}$ indicates that both expressions have the same probability law). We then adopt the parametrization $k=0$,$S(\alpha,\beta,\gamma,\delta;0)$
for which the random variable $X$ given by
\begin{flalign}
  X \overset{\delta}{=}
 \begin{cases}
  \gamma(Z-\beta \tan(\frac{\pi \alpha}{2}))+\delta,  \ & \alpha \neq 1 \\
  \gamma Z + \delta, \ & \alpha = 1
\end{cases},
\end{flalign}
has characteristic function
\begin{flalign}
 & S(\alpha,\beta,\gamma,\delta;0) \equiv E \exp(\imag u X) = \\ \nonumber
& \begin{cases}
  e^{\imag u \delta - \gamma^{\alpha}|u|^{\alpha}(1+\imag \beta(-1+|u \gamma|^{1-\alpha}){\rm sign}(u)  \tan(\frac{\pi \alpha}{2}))},  \ & \alpha \neq 1 \\
  e^{\imag u \delta - \frac{\gamma |u|(\pi+2\imag \beta \log(|u \gamma|)) {\rm sign}(u)}{\pi}}, \ & \alpha = 1
\end{cases},
\end{flalign}
where $Z=Z(\alpha,\beta)$ is defined as described in Eq.~(\ref{Zcf}), $\alpha
\in (0,2]$ is the index of stability or characteristic exponent, $\beta \in
[-1,1]$ the skewness parameter, $\gamma >0$ the scale parameter and $\delta \in \bm{R}$ location parameter. Two important special cases are the Gaussian
distribution with $\alpha =2$ and the Cauchy-Lorentz with $\alpha =1$ which are the only ones with an explicit closed formula. More details on the L\'evy
stable distribution can be found in \cite{Nolan}.

\begin{figure}[h]
\begin{center}
\includegraphics[width=8cm]{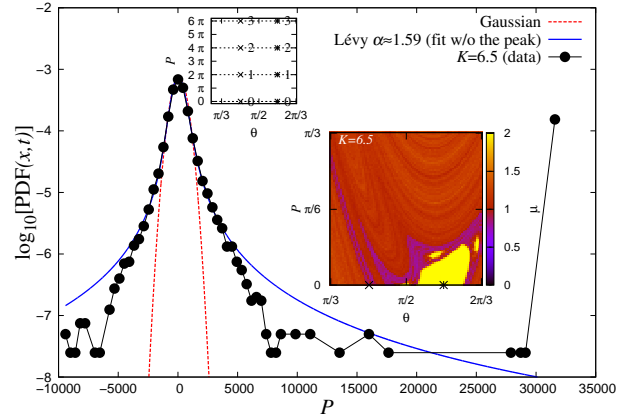}
\caption{\label{figLevyK} (Color online) The distribution of the momenta $P$
(black filled circles fitted by the L\'evy stable distribution for $K=6.5$. The best fit parameters parameters are found to be
$(\alpha,\beta,\gamma,\delta)\approx(1.59,0.164,3.7,83.63)$ while the ensemble
has $\approx$100000 $(314\times 314)$ mixed initial conditions (shown in the
inset), i.e. trajectories that are transported ballistically by the
\emph{unstable} accelerator mode around $(\theta_0,P)=(1.31179, 0)$ ($\times$)
and more evidently around the \emph{stable} one at $(\theta_1,P)=(\pi-\theta_0,0)$ ($\ast$), together with chaotic ones in their neighborhood after $n=5000$. In the inset color-map we show the diffusion exponent $\mu$, calculated by the process described in text, for a grid of cells on the subspace $(\theta,P) \in [\pi/3, 2\pi/3]\times[0,\pi/3]$ of the phase space. In yellow (light gray in b/w) color we present areas which correspond to ballistic motion due to the two accelerator modes pointed by the symbols $(\times)$ and $(\ast)$ respectively. Those initial conditions being inside ``pure'' chaotic regime diffuse normally with $\mu \approx 1$. In the edges of the accelerator mode areas where $\mu \approx 0.8$ one may find sticky subdiffusive transport (`belt''-like darker zone). The ballistic transport on the cylindrical phase space of the two above mentioned accelerator modes positioned initially at $(\theta_{0,1},0)$ and boosted by $\delta P = 2 \pi$ at every successive kick is shown in the small inset panel.}
\end{center}
\end{figure}

Let us now see how one may utilize the L\'evy stable distribution in order to
capture the effect of the anomalous diffusion due to the accelerator modes. In
Fig.~\ref{figLevyK}
we depict the stable L\'evy distribution with parameters
$(\alpha,\beta,\gamma,\delta)\approx(1.59,0.164,3.7,83.63)$ for $K=6.5$. In the inset color-map we show the exact grid of mixed initial conditions (i.e.
covering regular and chaotic regions) used ($314\times 314\approx$100000). This particular cell contains: (i) trajectories that are transported ballistically by the \emph{unstable} accelerator mode around $(\theta_0,P)=(1.31179, 0)$ ($\times$) and  (ii) more evidently around the \emph{stable} one at $(\theta_1,P)=(\pi-\theta_0, 0)$ ($\ast$) together with (iii) chaotic ones in their vicinity after $n=5000$. The fit for the L\'evy distribution was performed by excluding the last peak in the positive large momenta due to the ballistic transport by the accelerator mode. Generally speaking, one should expect two peaks in the distribution of the diffusive variable $P$ (in the positive and negative direction) depending on the choice of the ensemble of initial conditions. The color-plot inset panel shows the diffusion exponent $\mu$ for a grid of cells on the subspace $(\theta,P) \in [\pi/3, 2\pi/3]\times[0,\pi/3]$ of the phase space. The yellow (light gray in
black and white) color areas are related to ballistic motion due to the two
accelerator modes pointed by the symbols $(\times)$ and $(\ast)$ respectively.
Those initial conditions lying inside the fully chaotic regime diffuse normally with $\mu \approx 1$. Moreover, there is a ``belt''-like darker zone in the
edges of the accelerator mode area with $\mu \approx 0.8$ which indicates a
``sticky'' subdiffusive transport. In the small inset panel, one may see how
the ballistic transport takes place on the cylindrical phase space of the two
above mentioned accelerator modes positioned initially at $(\theta_{0,1},0)$
and boosted by $\delta P = 2 \pi$ at every (only four here) successive kick.
Furthermore, let us point out that the three yellow (or light gray in black and white) areas, surrounding the stable accelerator mode $(\ast)$, are
ballistically transported as well due to accelerator mode of higher period.

\begin{figure}[h]
\begin{center}
\includegraphics[width=8cm]{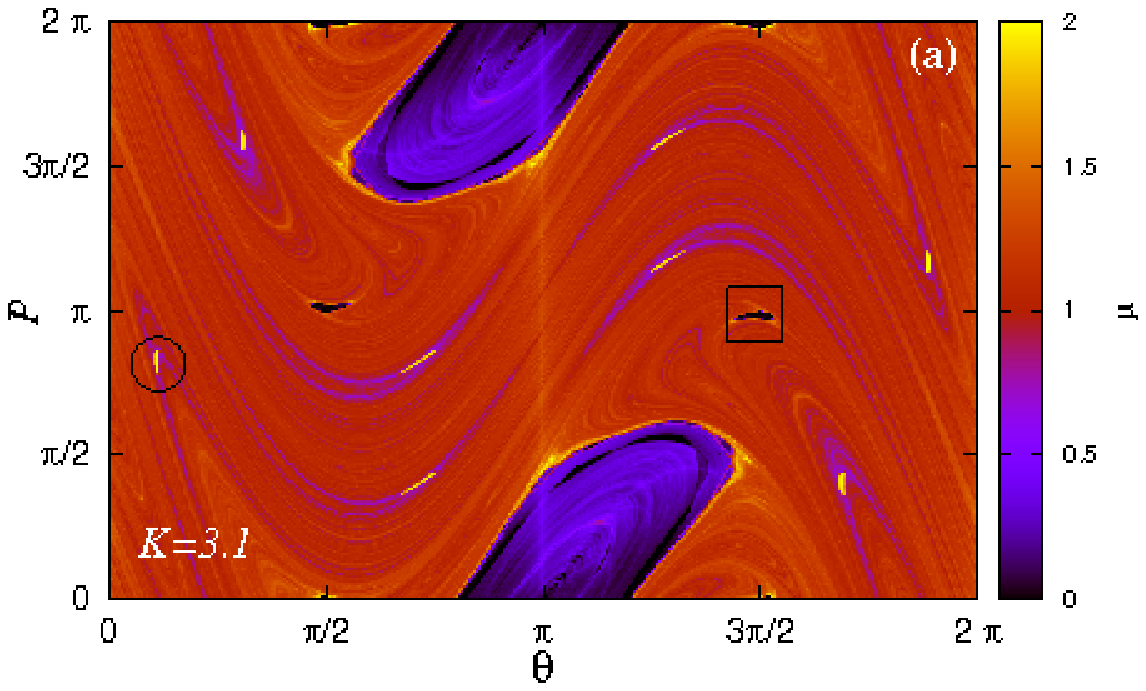}\\
\includegraphics[width=8cm]{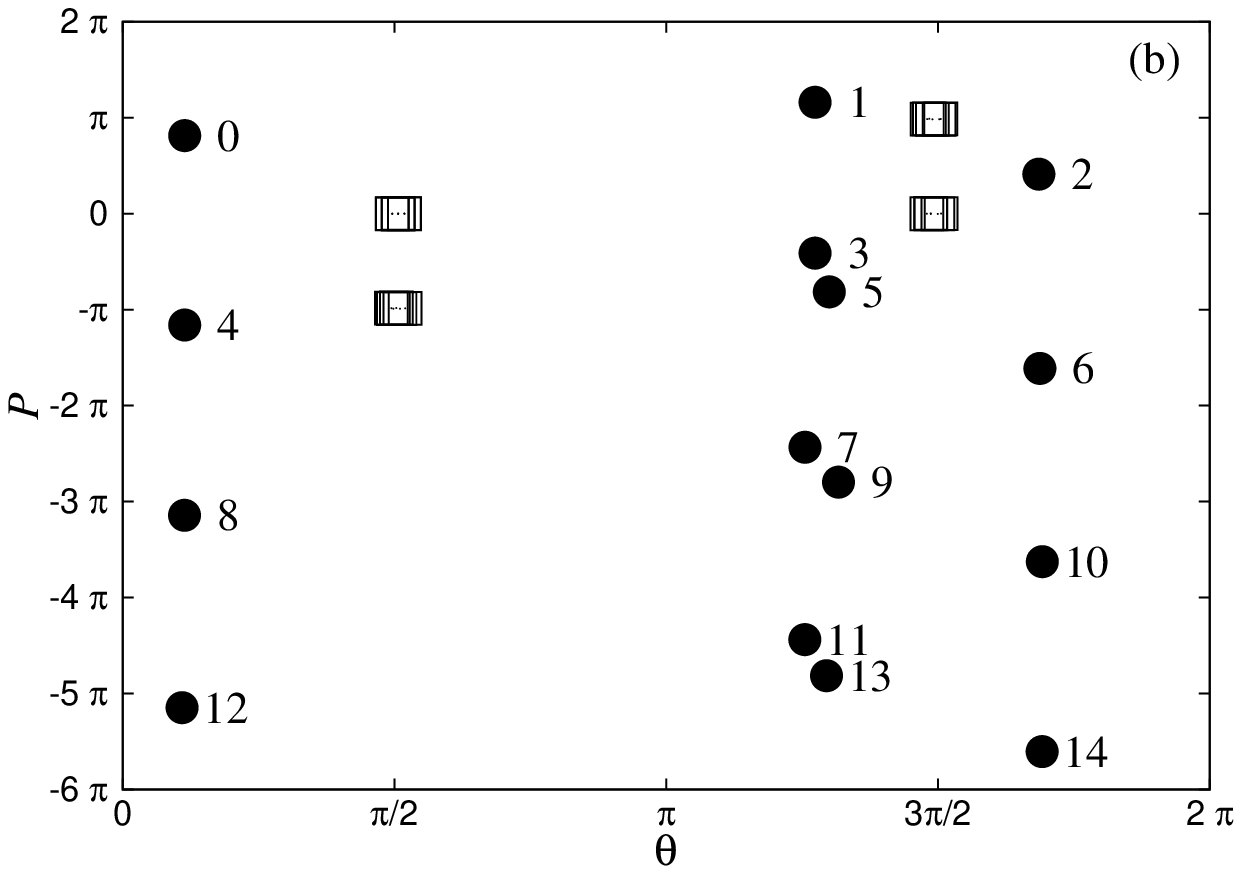}
\caption{\label{fig3} (Color online) (a) the diffusion exponent $\mu$ for
$K=3.1$ and for $50 \times 50$ initial conditions on a $500\times 500$ cell
grid of the entire phase space $(\theta,P) \in [0,2\pi] \times [0,2\pi]$
calculated after $n=5000$ iterations. (b) Two examples (marked with a square
and circle in panel a) following different diffusion processes: a trajectory
transported ballistically (with $P<0$) by the effect of an accelerator mode of
period 4 (circle) and one oscillating between islands of stability (square) of
period 4 too.}
\end{center}
\end{figure}

Color maps like the one presented in the inset of Fig.~\ref{figLevyK} can be
used for the (numerical) detection of accelerator modes in the phase space. In
Fig.~\ref{fig3}
we show how one can distinguish efficiently the regions with accelerator modes
from the islands of stability with the use of diffusion exponent $\mu$
color-plots. Let us first consider again a grid of $500\times 500$ cells (on
the entire phase space $(\theta,P) \in [0,2\pi]\times [0,2\pi]$) with $50
\times 50$ initial conditions in each and evolve all of them together for
$n=5000$ iterations. Then, for each ensemble of each cell separately, we
calculate numerically the diffusion exponent $\mu$ and append a color to this
cell (color bar of Fig.~\ref{fig3}(a)) that \emph{characterizes} the different
kind of diffusion of this small area. Depending on the location of each
ensemble and the color assigned to the cell, one may expect to find: (i)
\textit{ normal diffusion} ($\mu = 1$) inside chaotic regimes without the
presence of accelerator modes, (ii) \textit{subdiffusion} ($0< \mu < 1$) inside islands of stability, (iii) \textit{superdiffusion} ($1 < \mu < 2$) inside
chaotic regimes with the presence of accelerator modes in the phase space and
(iv) \textit{ballistic transport} ($\mu \approx 2$) inside and in the very
close vicinity of accelerator modes.

Note that the stable regions around $(\pi/2,0)$, $(3\pi/2,0)$, $(\pi/2,\pi)$,
$(3\pi/2,\pi)$ and $(\pi/2,2\pi)$, $(3\pi/2,2\pi)$ are correctly identified as
islands of stability since their diffusion exponent $\mu$ is smaller than 1.
The remaining tiny yellow areas correspond to stable higher period accelerator
modes with $\mu \approx 2$ (see \cite{MR2014} for more on the relation between
the diffusion process and the stability of the different regimes of the phase
space). Let us stress that following the above procedure, one manages to locate the accelerator modes not only of period 1 but also of higher periods 2,3,4,... The latter ones  are in general hard to be calculated analytically. Finally,
in Fig.~\ref{fig3}(b) we show two examples (marked with a square and a circle
in panel a) which present a rather different diffusion processes, i.e., for a
trajectory oscillating between islands of stability (squares) and for one
transported ballistically (with $P<0$) by the effect of an accelerator mode
(circles). Let us notice here that the period of both is 4 (when projected on
the $(\theta,P) \in [0,2\pi]$), as it can be seen when we iterate them for a
few steps.

\section{Summary}\label{sec:summary}

In this short paper we tried to give a small flavor of the accelerator modes'
impact in the diffusion process in the phase space of a system. These effects
are not limited only to the standard map but also to other models that can have such dynamical features. One of the main results is the link between the
superdiffusive transport and the existence of the accelerator modes for certain control $K$ parameter interval values. The interval size as well as the intensity (i.e. the deviation from normal diffusion expressed by the deviation of the diffusion exponent $\mu$ from $\mu=1$) is found to decrease as a power law, as $K$ increases (see Fig.~\ref{figKvsmu}). The effect of these modes seems to be strong even when their relative size in the phase space is small and even when one considers averages of initial conditions taken over the whole phase space. In the case where $K> 2\pi$ the accelerator modes are in general of period one, while for $K<2\pi$ they may be of higher period, and their influence on $\mu$ decays with $K$ much faster than for period one accelerator modes.

Looking into these resonances of the phase space even further, we tried to use
an adequate probability distribution function of the momenta $P$ that would
allow us to distinguish between normal and anomalous diffusion for different
ensembles of initial conditions. It turns out that the stable L\'evy
distribution and its $\alpha$ parameter can serve for this goal quite well,
despite the fact that this distribution is mainly introduced and broadly used
to capture random (L\'evy) walks in the literature. Here, we showed only one
exemplary case for $K=6.5$ (see Fig.~\ref{figLevyK}) where the $\alpha$-L\'evy
parameter was found to be $\approx 1.59$ while the $\mu$ diffusion exponent
lies on a range of values depending on the location of the sub-cell of initial
conditions (inset of Fig.~\ref{figLevyK}).

Finally, by considering a fine grid in the phase space and then taking
ensembles of many initial conditions for the averages inside the small cells, we managed to produce diffusion ``chart-map'' encoding their different
(numerically estimated) $\mu$ value as shown for example in Fig.~\ref{fig3}.
With this kind of plots one may detect and separate accelerator modes ($\mu
\approx 2$) of different periods from even small islands of stability
$(0<\mu<1)$, or chaotic regions ($\mu \approx 1$), or ``sticky'' intermediate
regimes.

We should mention that another important method to study the structure of the phase portrait rests upon calculating the Lyapunov exponents in order to distinguish the regular from the chaotic regions in the phase space, and also allows us to quantify the degree of chaos by the value of the Lyapunov exponents \cite{Skokos2010} at the sufficiently densely and uniformly chosen initial conditions. However, the calculation of the Lyapunov exponents can be quite laborious task, while the so-called GALI method, a generalization
of the original SALI method, introduced and developed by Skokos et al
\cite{Skokos2001,Skokos2004,Skokos2007,Skokos2008}, and recently reviewed by Skokos and Manos in reference \cite{SM2014}, is an extremely efficient method, which distinguishes quickly the regular from chaotic orbits, although it does not quantify the degree of chaos to such an extent as the Lyapunov exponents.
This method has been used in our main work \cite{MR2014} and it was found that there is excellent agreement between the $\mu$-landscape and the GALI-landscape in the phase space of the standard map. We believe that our approach and the method can be applied in other smooth Hamiltonian dynamical systems and mappings.


\end{document}